\documentclass[onecolumn]{aastex63}
\usepackage{rotating} 
\usepackage{amsmath,amssymb,bbold,mathrsfs}
\usepackage{subfigure}
\usepackage{bm}

\makeatletter
\let\frontmatter@title@above=\relax
\makeatother

\shorttitle{The CLASP2 near ultraviolet \ion{Fe}{2} lines}
\shortauthors{Afonso Delgado et al.}

\graphicspath{{./}{figures/}}

\begin{document}

\title{Magnetic field information in the 
near-ultraviolet\\ \ion{Fe}{2} lines of the CLASP2 space experiment}

\email{dafonso@iac.es}

\author{David Afonso Delgado}
\affil{Instituto de Astrofísica de Canarias, E-38205 La Laguna, Tenerife, Spain}
\affil{Universidad de La Laguna, Dept. Astrofísica, E-38206, La Laguna, Tenerife, Spain}
\author{Tanaus\'u del Pino Alem\'an}
\affil{Instituto de Astrofísica de Canarias, E-38205 La Laguna, Tenerife, Spain}
\affil{Universidad de La Laguna, Dept. Astrofísica, E-38206, La Laguna, Tenerife, Spain}
\author{Javier Trujillo Bueno}
\affil{Instituto de Astrofísica de Canarias, E-38205 La Laguna, Tenerife, Spain}
\affil{Universidad de La Laguna, Dept. Astrofísica, E-38206, La Laguna, Tenerife, Spain}
\affil{Consejo Superior de Investigaciones Científicas, Spain}

\begin{abstract}

We investigate theoretically the circular polarization signals induced by the Zeeman effect in
the \ion{Fe}{2} lines of the $279.3$~-~$280.7$~nm 
spectral range of the CLASP2 space experiment and their suitability to infer solar magnetic fields.
To this end, we use a comprehensive \ion{Fe}{2} atomic model 
to solve the problem of the generation and transfer of polarized 
radiation in semi-empirical models of the solar
atmosphere, comparing the region of formation of the \ion{Fe}{2} spectral lines with those of the
\ion{Mg}{2} h and k and the \ion{Mn}{1} resonance lines. 
These are present in the same near ultraviolet (near-UV) spectral region and allowed 
the mapping of the longitudinal component of the magnetic field ($B_{\rm L}$) through several
layers of the solar chromosphere in an active region plage. 
We compare our synthetic intensity profiles with observations from the IRIS and CLASP2 missions, 
proving the suitability of our model atom to characterize these \ion{Fe}{2} spectral lines.
The CLASP2 observations show two \ion{Fe}{2} spectral lines at 279.79 and 280.66~nm 
with significant circular polarization signals. We demonstrate the suitability 
of the Weak-Field Approximation (WFA) applied to the Stokes $I$ and $V$ profiles 
of these \ion{Fe}{2} lines to infer $B_{\rm L}$ in the plage atmosphere.
We conclude that the near-UV spectral region of CLASP2 allows to determine $B_{\rm L}$
from the upper photosphere to the top of the chromosphere of active region plages.

\end{abstract}



\section{Introduction} \label{sec:intro}

Understanding the structure and evolution of the magnetic field throughout the
solar atmosphere is key to understanding the transfer of energy from the bottom 
of the photosphere
to the chromosphere and to the million-degree corona. It is thus clear that observations and
diagnostic tools allowing for a reliable and simultaneous inference of
the magnetic field in different layers of the solar atmosphere, from the lower photosphere to
the top of the chromosphere, just below the transition region to the corona, are necessary to
improve our understanding of the physical processes taking place in the atmosphere of our
closest star \citep[e.g.,][]{TrujilloBueno2022}.

Spectral lines encode information about the physical conditions of the plasma in their region
of formation in the solar atmosphere. Many of the lines that form in the outer layers
of the solar chromosphere are located in the UV region of the spectrum, which gives
this spectral region significant potential for magnetic-field diagnostics across the solar atmosphere.
For a recent review on the physics and diagnostic potential of UV spectropolarimetry, 
see \cite{TrujilloBueno2017}.
In particular, two of the strongest and brightest spectral lines 
in the near-UV spectrum, the \ion{Mg}{2} h and k
resonant doublet, are located at 280.35 and 279.64~nm, respectively. The line center of
these spectral lines originate just below the transition region, while their wings form in deeper
layers between the lower 
and middle chromosphere \citep[e.g.,][]{Leenaarts2013a}. Consequently, these
spectral lines are among the most important for the study of the solar chromosphere.

Due to the significant opacity of the Earth's atmosphere at UV wavelengths, 
these spectral lines can
only be observed from space. The first spectropolarimetric data on the
\ion{Mg}{2} h and k lines were obtained by the Ultraviolet Spectrometer and Polarimeter
\cite[UVSP,][]{Calvert1979, Woodgate1980} onboard the Solar Maximum Mission \cite[SMM,][]{Bohlin1980},
which indicated the existence of significant scattering polarization signals 
in the far wings of these lines \citep{Hence1987,MansoSainz2019}.
Since 2013, their intensity spectrum has been systematically observed with the Interface Region Imaging
Spectrograph mission \cite[IRIS,][]{dePontieu2014}, whose results have improved our
knowledge about the thermal and dynamical structure of the upper layers of the solar chromosphere
\cite[see the review by][]{dePontieu2021}.

Motivated by several theoretical studies on the polarization of the \ion{Mg}{2} h and k lines around 280~nm
(\citealt{Belluzzi-TrujilloBueno2012,DelPinoAleman2016,AlsinaBallester2016}, and \citealt{delPinoAleman2020}),
the Chromospheric LAyer SpectroPolarimeter \cite[CLASP2,][]{Song2022} 
suborbital mission was launched
on April 11 2019, achieving unprecedented observations of 
the four Stokes parameters of this near-UV spectral region in an active region plage
close to the solar disk center and in a quiet Sun region near the limb. 
The analysis of the quiet Sun spectral
profiles confirmed the theoretical predictions on the scattering 
polarization signals \citep{Rachmeler2022,Ishikawa2023}.
The application of the Weak-Field Approximation (WFA) to the circular polarization profiles observed in the plage region allowed for
the mapping of the longitudinal component of the magnetic field ($B_{\rm L}$) across different layers of
the solar chromosphere \citep{Ishikawa2021a}, extended to the photosphere with simultaneous observations
by Hinode/SOT-SP \citep{Hinode}. Later, the application of the Tenerife Inversion Code \cite[TIC,][]{Li2022} 
to the CLASP2 Stokes $I$ and $V$ profiles observed in the plage target  
allowed for the inference of a stratified model atmosphere, including
its thermal, dynamic, and magnetic structure \citep{Li2023}.

Even though the analysis of the CLASP2 data has been limited, 
up to now, to the \ion{Mg}{2} and \ion{Mn}{1}
lines, these observations show several other spectral lines with significant circular polarization signals.
Among these lines, two \ion{Fe}{2} spectral lines stand out: a relatively weak emission line at 279.79~nm and
an absorption line at 280.66~nm, just at the red edge of the CLASP2 spectral window.
Recent investigations have suggested that the \ion{Fe}{2} lines located 
blueward of the 
CLASP2 spectral region are of complementary interest for diagnosing 
the magnetism of the solar chromosphere \citep[see][]{Judge2021,Judge2022}.
A quantitative study of the polarization signals and magnetic sensitivity of the \ion{Fe}{2} spectral lines
in the near-UV region of the solar spectrum has been recently published by \cite{Afonso2023b},
but the \ion{Fe}{2} lines located in the CLASP2 spectral region were not included in their work.

The pioneering results of the CLASP2 mission have proven the capabilities
of the \ion{Mg}{2} h and k spectral lines to study the magnetic 
field in the upper layers of the chromosphere.
There has been a growing interest in the planning and development of space missions 
to perform full-Stokes observations of this spectral window, such as the Chromospheric Magnetism
Explorer \cite[CMEx,][]{CMEx}. The analysis of the lines from species other than \ion{Mg}{2} found in this spectral
region is critical to take full advantage of the possibilities that future space missions bring
to map the evolution of the magnetic field throughout the solar atmosphere.

In this paper we study the formation of the \ion{Fe}{2} spectral 
lines in the CLASP2 spectral window, as well as
their suitability to infer $B_{\rm L}$ and how they complement 
the magnetic field information that can be inferred
from the \ion{Mg}{2} and \ion{Mn}{1} lines in the same spectral window \citep[see][]{Ishikawa2021a}. In
Section~\ref{sec:problem} we describe the atomic 
models and our strategy to solve the radiation transfer (RT)
problem. In Section~\ref{sec:StkI} we analyze the 
emergent synthetic intensity profiles, studying the suitability
of our model atom by comparing with observations 
at different lines of sight (LOS).
In Section~\ref{sec:results} we study the formation of 
the circular polarization profiles of the \ion{Fe}{2} lines
at 279.79 and 280.66~nm, and the suitability of the WFA 
to infer $B_{\rm L}$. We then apply the WFA to the CLASP2
data of these \ion{Fe}{2} lines. Finally, in Section~\ref{sec:conclusions} we present our conclusions.


\section{Formulation of the problem} \label{sec:problem}

Despite the significant number density of \ion{Fe}{2} in the solar atmosphere, producing some of the
strongest spectral lines in the near-UV solar spectrum (e.g., the resonant multiplet at 260 nm), 
the \ion{Fe}{2} lines in the 279.3~-~280.7~nm 
spectral window are relatively weak (see Table~\ref{tab:atomic}
for their atomic data). In particular, both \ion{Fe}{2} lines at 279.79 and 280.66~nm are intercombination
lines. Because they form in the scattering wings of the \ion{Mg}{2} h and k doublet,
including the latter is necessary for an accurate modeling of these \ion{Fe}{2} lines. Moreover, we also
include an atomic model for the \ion{Mn}{1} resonant  triplet, also present in this spectral window.

As it is the case with most of the atomic lines located in the near-UV solar spectrum, the modeling of
their intensity and polarization requires to account for strong departures from local thermodynamic equilibrium 
\cite[LTE; e.g.,][]{MihalasBook}. In particular, one must jointly solve the statistical equilibrium equations (SEE) describing
the populations and coherence between atomic sublevels, and the RT equations, 
describing the propagation of the electromagnetic radiation as it travels through the atmospheric plasma 
\cite[e.g.,][]{LL04}. The \ion{Mg}{2} h and k
spectral lines show significant partial frequency redistribution (PRD) effects, and the modeling of their
polarization requires accounting for $J$-state interference \citep{Belluzzi-TrujilloBueno2012,Belluzzi-TrujilloBueno2014,DelPinoAleman2016}.
Therefore, we account for $J$-state interference in our \ion{Mg}{2} atom model \citep[multi-term model atom; see \S7.6 in][]{LL04}, 
but we neglect it in the cases of the \ion{Fe}{2} and \ion{Mn}{1} atom models \citep[multi-level model atom; see \S7.2 in][]{LL04}.
We solve the RT problem accounting for all these physical ingredients with the HanleRT code
\citep{DelPinoAleman2016,delPinoAleman2020}. 

Our magnesium model atom comprises three \ion{Mg}{2} terms, including the resonant doublet with PRD and the
UV subordinate triplet with complete frequency redistribution (CRD), and the ground term of \ion{Mg}{3}.
This atomic model is the same used by \cite{delPinoAleman2020} and \cite{Afonso2023}.  Our atomic model for iron is the one described in
\cite{Afonso2023b}, to which we have added the atomic transition corresponding to the line at 279.79~nm,
between the levels a$^4$F$_{3/2}$~-~z$^6$D$^\circ_{3/2}$ (the upper level is in the
upper term of the resonant multiplet at 260~nm). The oscillator strength of this transition was
taken from \cite{KuruczDB}. 
Moreover,
we determine the rate of inelastic collisions with free electrons and the photoionization cross-section
by applying the approximations of \cite{vanRegemorter1963}, \cite{Bely1970}, and \cite{AllenBook1963}. We
adjust the collisional rates through an ad-hoc factor in order to fit the line core intensity flux
of the \ion{Fe}{2} resonant transitions between 258.6 and 260.25~nm in the quiet Sun C model of
\cite{Fontenla1993} to the observed flux of the solar analog $\alpha$ Cen A \citep[see][for further details]{Afonso2023b}.

An accurate modeling of the \ion{Mn}{1} resonant doublet requires accounting for the hyperfine
structure \citep[HFS, see][]{delPinoAleman2022}. The impact of the HFS is particularly relevant
regarding the circular polarization profiles. However, it is not critical
for estimating the region of formation of the lines. In this work we thus neglect the HFS, using
an atomic model with four \ion{Mn}{1} levels and the ground level of \ion{Mn}{2}, with the
resonant triplet in CRD.

Accounting for PRD and $J$-state interference in \ion{Mg}{2} and, at the same time, for the
large number of levels and transitions in the \ion{Fe}{2} model atom, makes the simultaneous
solution of the RT problem prohibitively expensive in terms of computing resources. Moreover,
solving the non-LTE problem for both \ion{Fe}{2} and \ion{Mg}{2} affects the ionization
balance of the latter (one of the reasons being the lack of \ion{Fe}{1} as a source of background
opacity in the far-UV). Therefore, we have opted for the following strategy:

\begin{enumerate}
    \item Calculate jointly the population balance for \ion{Mg}{2} and \ion{Mn}{1}.
    \item Calculate, independently, the population balance for \ion{Fe}{2}.
    \item Solve jointly the non-magnetic RT problem with the three atomic models by fixing the populations
          calculated in steps 1 and 2.
    \item For the magnetic field problem, use a reduced model of \ion{Fe}{2} with just five
          atomic levels (the upper and lower levels of the transitions at 279.79 and 280.66~nm, and the ground level of \ion{Fe}{3}), still fixing the populations to those calculated in steps 1 and 2.
\end{enumerate}

In this work we solve the RT problem in the 1D, plane-parallel, semi-empirical C and P atmospheric models described in
\citet[][hereafter FAL-C and FAL-P, respectively]{Fontenla1993}, representative of the quiet Sun and a
plage region.

\begin{deluxetable*}{cccccc} \label{tab:atomic}
\tablecaption{Atomic data and effective Land\'e factor for \ion{Fe}{2} atomic transitions in the spectral region of CLASP2.}

\tablehead{
\colhead{$\lambda$~[nm]} & \colhead{Transition} & \colhead{Levels energy} & \colhead{A$_{ul}$~[s$^{-1}$]} & \colhead{g$_{eff}$} }

\startdata
 279.471 &   a$^4$G$_{9/2}$ - z$^4$I$^\circ_{11/2}$  &  25 805 - 61 587    & $1.3\cdot10^7$   &  0.78$^{\rm a}$ \\
 279.787 &   a$^4$F$_{3/2}$ - z$^6$D$^\circ_{5/2}$   &\, 3 117 - 38 858    & $1.9\cdot10^4$   &  2.60$^{\rm b}$  \\
 280.012 &   b$^2$H$_{9/2}$ - y$^4$F$^\circ_{7/2}$   &  26 352 - 62 065    & $1.5\cdot10^7$   &  0.45$^{\rm b}$  \\
 280.485 &   a$^2$F$_{5/2}$ - x$ ^4$D$^\circ_{5/2}$  &  27 620 - 63 272    & $1.6\cdot10^6$   &  1.01$^{\rm b}$\\
 280.661 &   a$^2$F$_{7/2}$ - x$^4$D$^\circ_{7/2}$   &  27 314 - 62 945    & $3.2\cdot10^6$   &  1.38$^{\rm b}$\\
\enddata
$^{\rm a}$LS coupling, accounting for the upper level coupling between $^4$H$^\circ_{11/2}$ and $^4$G$^\circ_{11/2}$

$^{\rm b}$experimental \citep{NIST_ASD}
\end{deluxetable*}


\section{Intensity Profiles} \label{sec:StkI}

\begin{figure*}[htp!]
\plotone{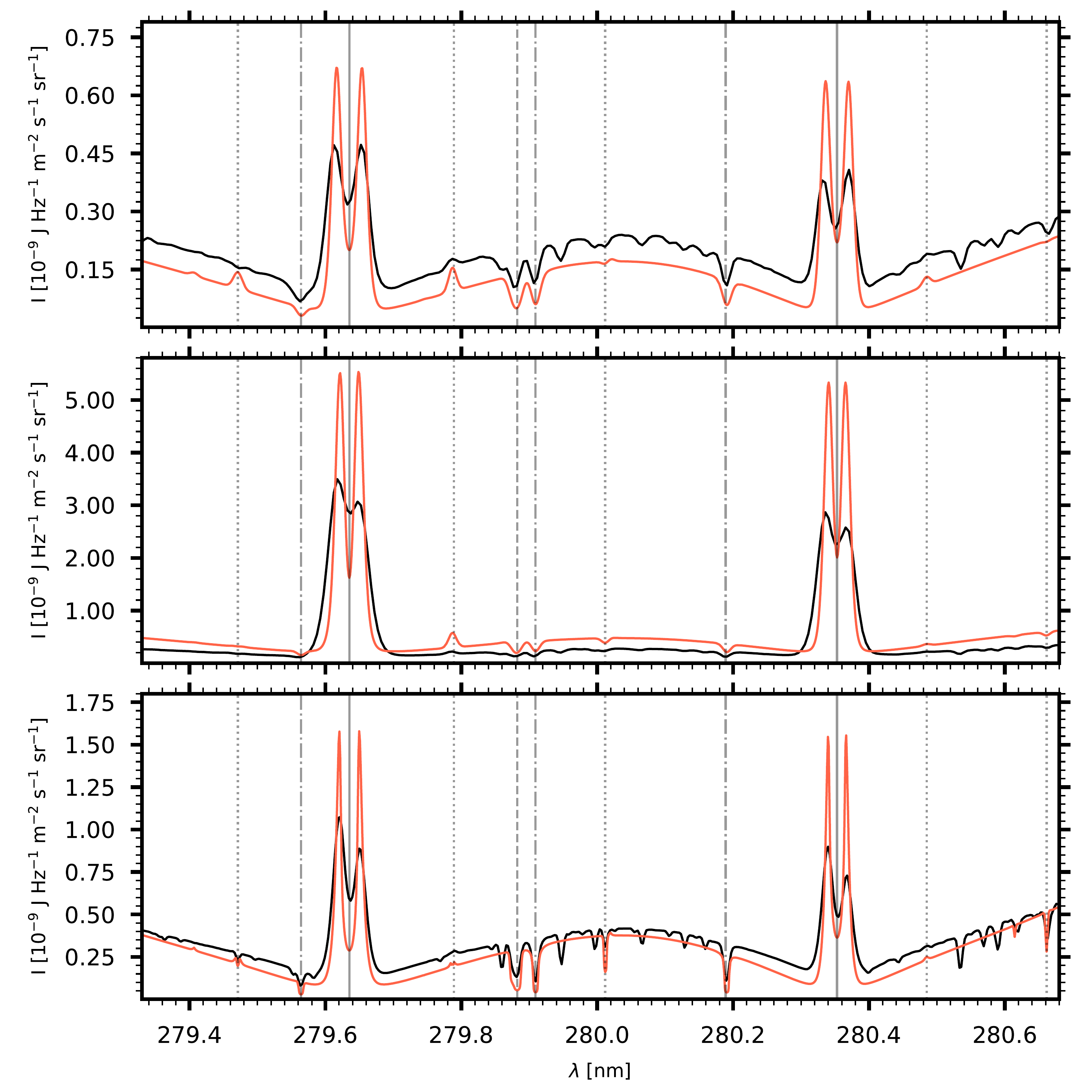}
\caption{Emergent intensity profile for the spectral region observed by the CLASP2 mission for
an LOS close to the solar limb ($\mu=0.1$; upper panel), 
for a pixel located in the active solar plage observed by CLASP2 ($\mu=0.7$, middle panel), 
and for an LOS at the solar disk center
($\mu=1$; lower panel). The red solid curves in upper and lower panels show our calculation in the FAL-C  model 
atmosphere and in the middle panel it shows our calculation in the FAL-P model atmosphere.
The black solid curve in the top (middle) panel is an average over 2 arcsec of the CLASP2 slit limb (plage)
target observation. The black solid curve in the bottom panel is the average over all the field of view
in a quiet Sun region observed by IRIS on October 4 2022. The vertical gray lines indicate the
positions of the spectral lines:
\ion{Mg}{2} h and k (solid), \ion{Mg}{2} subordinate lines (dashed), \ion{Mn}{1} resonant triplet
(dashed-dotted), and \ion{Fe}{2} lines (dotted).}
\label{fig:intensity}
\end{figure*}

We compare, qualitatively, the synthetic intensity profiles with available observations
(see Fig.~\ref{fig:intensity}). Within the CLASP2 spectral window we can find five
\ion{Fe}{2} spectral lines showing significant absorption or emission signals in both
the theoretical and observed intensity profiles (indicated with gray-dotted lines in the figure).
The upper panel of Fig.~\ref{fig:intensity} shows the theoretical profile for an LOS
with $\mu=\cos{\theta}=0.1$ (with $\theta$ the heliocentric angle of the observed point) 
together with the CLASP2 observation
of a quiet Sun region 
close to the solar limb after averaging over 2 arcsec along the slit. The lower panel shows a comparison of the theoretical profiles for
an LOS with $\mu=1.0$ together with the average of all the profiles contained in a whole map of an IRIS observation of the quiet Sun
at the solar disk center from October 4 2022. 
Despite the limitations of using semi-empirical and static atmospheric models, 
the synthetic spectral lines show a qualitative behavior close to that of the observations.
One exception to this is the \ion{Fe}{2} line at 279.47~nm which seems
to be not accurately represented by our atomic model. 
In the observations this spectral line is always in absorption while in our calculations
it shows clear emission features both at disk-center and close to the limb.

There is a notable difference between synthetic and observed profiles in the \ion{Mg}{2} h
and k wings (see Fig.\ref{fig:intensity}). This disagreement is due to the different temperature in
the formation region of the wings (upper photosphere) between the semi-empirical models and the actual
atmospheric plasma. In the quiet-Sun (plage) observations, this temperature is lower
(higher) than in the FAL-C (FAL-P) model. Note, however, that this difference in the quiet Sun observations
is significantly diminished when comparing with the disk center observation from IRIS (bottom panel of
Fig.\ref{fig:intensity}). In this case we average a larger number of profiles and, consequently,
the semi-empirical ``average'' quiet Sun model FAL-C is much closer to the average of the quiet Sun
observation.

The \ion{Fe}{2} lines at 280.01 and 280.66~nm show significant absorption for an LOS at disk center
and a relatively weaker absorption for an LOS toward the limb, for both theoretical and observed 
profiles. The \ion{Fe}{2} lines at 279.79 and 280.48~nm show a relatively weak emission at disk 
center, which becomes more significant for LOS closer to the limb. We emphasize, in particular,
that the model successfully reproduces the behavior for different LOS for the \ion{Fe}{2}
line at 279.79~nm, almost imperceptible for the disk center LOS and with a clear emission
for an LOS close to the limb.

The CLASP2 observations in the plage region also show a significant emission feature in the
\ion{Fe}{2} line at 279.79~nm \citep[see][]{Ishikawa2021a}, despite the location of the plage being
close to disk center (between $\mu=0.8$ and $0.6$). A spectral synthesis in the FAL-P model,
representative of a solar plage, also shows a remarkable emission
(see middle panel in Fig.~\ref{fig:intensity}).

These results support the suitability of the \ion{Fe}{2} atomic 
model presented in \cite{Afonso2023b} for the present research.


\section{Circular Polarization Profiles} \label{sec:results}

In this section we study the formation of the circular polarization profiles of the
\ion{Fe}{2} lines in the CLASP2 spectral window. To this end, we impose 
an ad-hoc exponential stratification of the magnetic field (see the right panel of Fig.~\ref{fig:wfa_theor}) in the FAL-C model atmosphere.
None of the \ion{Fe}{2} spectral lines mentioned in Sec.~\ref{sec:StkI} show any linear polarization signals, either in the CLASP2 observations or in our numerical modeling.
These lines show a quenching of the linear polarization in the
\ion{Mg}{2} h and k wings. 
This depolarization appears to be insensitive to artificial
changes of the rate of the depolarizing collisions of the \ion{Fe}{2} atomic model, indicating that the lines are just
depolarizing the \ion{Mg}{2} wings and do not show intrinsic polarization. Therefore,
from now on we focus on the circular polarization profiles.

\subsection{Theoretical Circular Polarization Profiles.} \label{sec:Results_Theor}

In the CLASP2 observations of an active region plage, only two of the \ion{Fe}{2} spectral
lines mentioned in Sec.~\ref{sec:StkI} show clear circular polarization signals. These are 
the \ion{Fe}{2} lines  at 279.79 and 280.66~nm. The \ion{Fe}{2} lines at 279.471 and
280.012~nm have relatively small effective Land\'e factors (0.78 and 0.45, respectively),
and their circular polarization signals may be below the noise level as shown in Fig.~\ref{fig:wfa}. The \ion{Fe}{2}
at 280.485~nm is very weak for an LOS close to disk center (the plage region
observed by CLASP2 is at $\mu\sim0.8$), which can explain the absence of clear circular
polarization signals. Therefore, in the following we focus on the \ion{Fe}{2} lines
at 279.79 and 280.66~nm.

\begin{figure*}[htp!]
\plotone{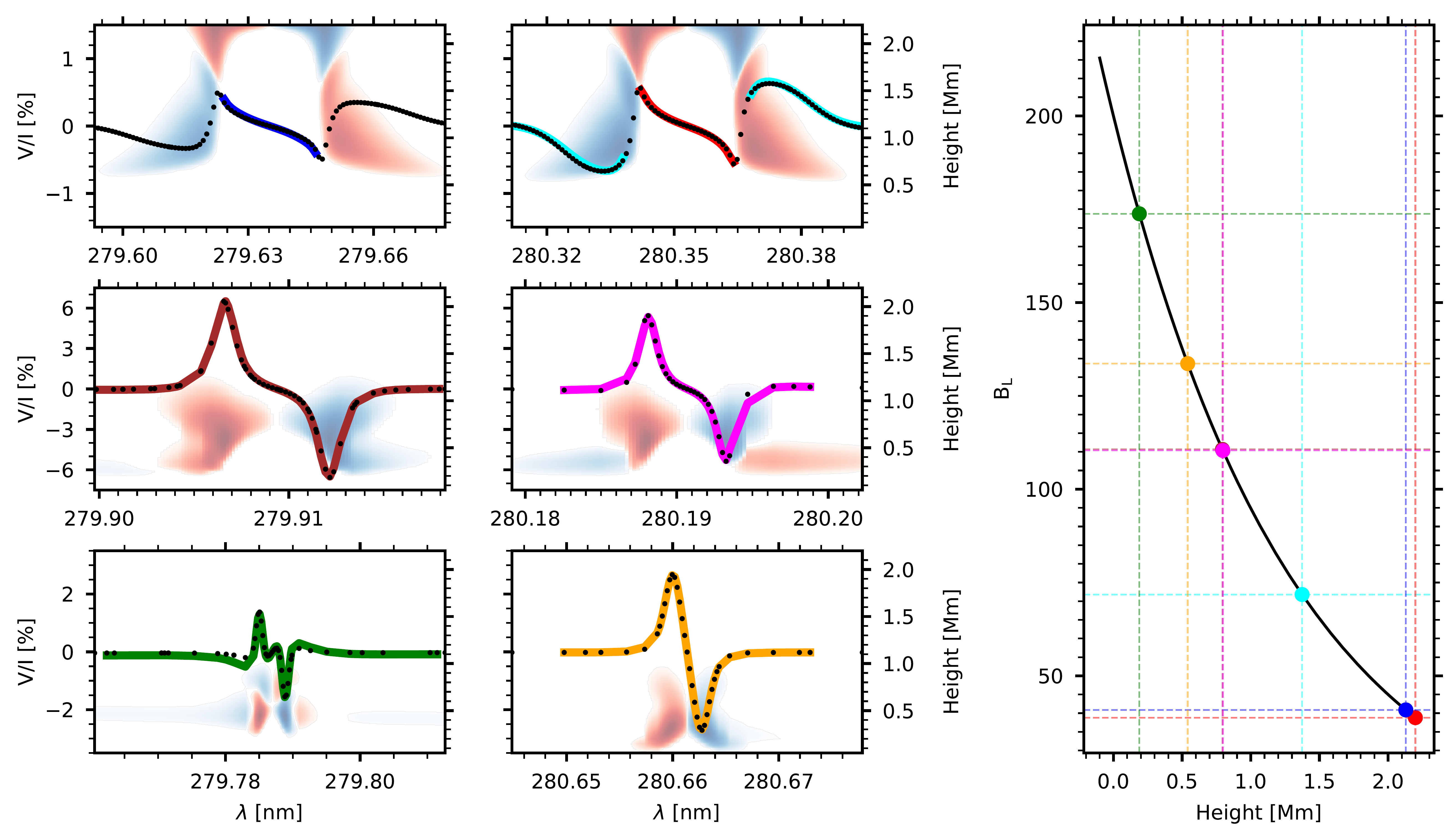} 
\caption{Emergent fractional circular polarization profiles $V/I$ (first and second
columns) for the \ion{Mg}{2} k and h lines (first row), \ion{Mn}{1} lines at 279.91
and 280.19~nm (second row), and \ion{Fe}{2} lines at 279.787 and 280.661~nm
(third row), for a disk center LOS, calculated in the FAL-C model with a vertical
magnetic field with an exponential stratification (see rightmost panel).
The colored background in the first two columns shows the logarithm of the
normalized contribution function for the circular polarization (with the red and
blue colors indicating positive and negative contributions, respectively).
The colored curves in the first and second columns show the fit with the WFA.
The outer lobes of the k line are excluded from the WFA fitting.
The colored circles in the rightmost panel show the $B_{\rm L}$ value from the fit with
the WFA of the circular polarization profile corresponding to the curve of the same
color, placed at its intersection with the $B_{\rm L}$ stratification of the model.
The colored solid lines help to quickly find the height and $B_{\rm L}$ values for each
colored circle in the intersection between each line and the coordinate axes. 
The brown and magenta lines and circles overlap in the
rightmost panel.
}
\label{fig:wfa_theor}
\end{figure*}

\begin{figure*}[htp!]
\plotone{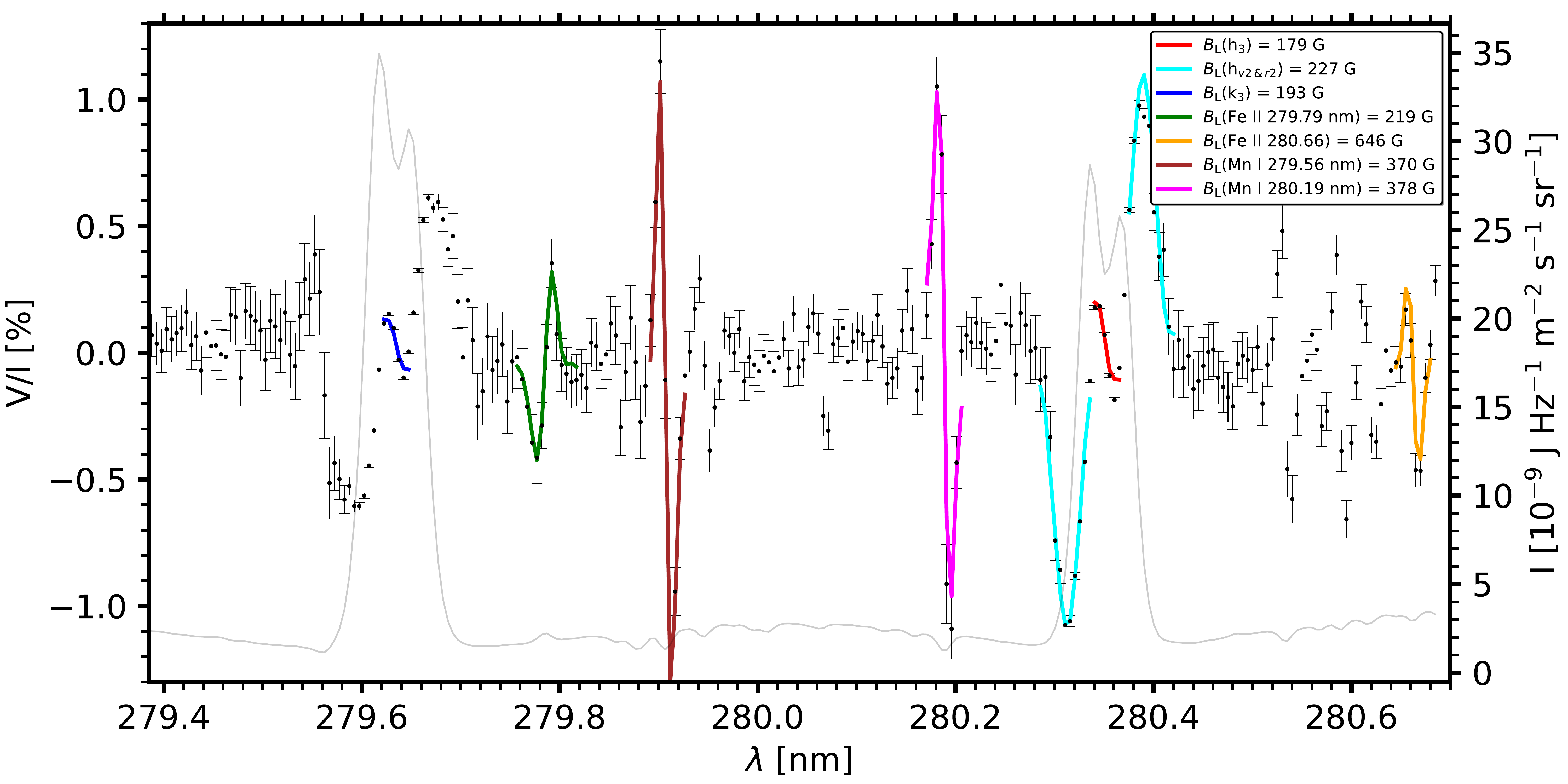} 
\caption{Fractional circular polarization $V/I$ in a selected pixel in the
CLASP2 plage observation (black dots, with error bars). The gray line
shows the corresponding intensity profile (see right axis). The colored
curves show the WFA fit for the seven spectral regions indicated in 
Fig.~\ref{fig:wfa_theor}, in the same colors: the inner and outer lobes of the
\ion{Mg}{2} h line (red and light blue, respectively), the inner lobes of the \ion{Mg}{2}
k line (blue), the \ion{Mn}{1} resonant lines (brown and magenta), and
the \ion{Fe}{2} lines at 279.79 and 280.66~nm (green and orange, respectively).
The $B_{\rm L}$ inferred for each of them is indicated in the top-right inset.
}
\label{fig:wfa}
\end{figure*}

To study in detail the circular polarization signals in these two lines, we
solve the RT problem in the FAL-C model atmosphere for a disk-center LOS as described in
Sec.~\ref{sec:problem}, assuming a vertical magnetic field with strength decreasing
exponentially with height, from 200~G in the bottom of the photosphere to 40~G just below the
transition region (see the right panel of Fig.~\ref{fig:wfa_theor}).

Under proper conditions, the WFA provides a fast method to estimate $B_{\rm L}$.
For it to be applicable, the Doppler width of the line
$\Delta\lambda_D = \frac{\lambda_{0}}{c} \sqrt{\frac{2kT}{m} + v_{\rm m}^2}$, (with $T$
and $v_{\rm m}$ the temperature and turbulent velocity, respectively, $c$ the speed of light, 
$k$ the Boltzmann constant, $m$ the mass of the atom, and $\lambda_0$ the line's wavelength), must be much larger than the
magnetic Zeeman splitting
($\Delta\lambda_B = 4.6686\cdot10^{-13}B\lambda_0^2$, with $B$ the magnetic field strength
in gauss and $\lambda_0$ in angstroms); i.e.

\begin{equation} \label{Eq:WFAcondition}
g_{\rm eff}\frac{\Delta\lambda_B}{\Delta\lambda_D} \ll 1.
\end{equation} 
When this condition is met, and $B_{\rm L}$ is constant across the line formation region,
the circular polarization profile is proportional to the wavelength 
derivative of the intensity
(e.g., \cite{LL04})
\begin{equation}\label{eq:wfa}
    V = -4.6686\cdot10^{-13}g_{\rm eff}\lambda_0^2B_{L}\frac{\partial I}{\partial \lambda}.
\end{equation}

The first and second columns of Fig.~\ref{fig:wfa_theor} show the circular
polarization of the \ion{Mg}{2} k and h lines (upper row), the \ion{Mn}{1} lines at
279.91 and 280.19~nm (middle row), and the \ion{Fe}{2} lines at 279.79 and 280.66~nm
(bottom row). The colored background shows the logarithm of the normalized
contribution function for Stokes $V$, which indicates the contribution of each region
in the model atmosphere (see the height in the right axis in each panel) to the
emergent circular polarization profiles. Red (blue) color indicates positive (negative) 
contributions to the circular polarization.

As shown in previous works, the contribution function indicates that the circular
polarization in the inner lobes of the \ion{Mg}{2} h and k lines form in the upper
chromosphere, while the circular polarization in the outer lobes forms in a relatively
extensive region in the middle chromosphere of the FAL-C model. The application of
the WFA independently to each of these spectral regions allows us to
fit the circular polarization profiles of all the selected spectral lines.
We have excluded the outer lobes
of the \ion{Mg}{2} k line in this work because the blue outer lobe is blended with a
\ion{Mn}{1} line, impacting its circular polarization signal. The $B_{\rm L}$ values inferred
with the WFA are found in the model atmosphere at those heights with the largest values
of the contribution function (compare the contribution function in the first and
second columns of Fig.~\ref{fig:wfa_theor} with the colored dots in the rightmost panel).

The region of formation of the circular polarization of the \ion{Mn}{1} resonant
lines is below that of the outer lobes of the \ion{Mg}{2} h and k, thus between the
lower and middle chromosphere in FAL-C. 

Moreover, the WFA
can be applied to the observations using the effective Land\'e factor calculated
assuming LS coupling without HFS \citep{delPinoAleman2022}. All the results shown
here regarding the \ion{Mg}{2} and \ion{Mn}{1} lines in this spectral window agree
with previous results
(\citealt{delPinoAleman2020}, \citeyear{delPinoAleman2022}; \citealt{Ishikawa2021a}).

The \ion{Fe}{2} line at 279.79~nm shows a circular polarization profile with
four lobes, due to what seems to be a self-absorption feature in the center of
the line (see the lower panel of Fig.~\ref{fig:intensity}). The inner lobes resulting from the
numerical calculation cannot be observed with the CLASP2 spectral sampling
($\sim$0.01~nm), and thus the CLASP2 observations show a two-lobes
circular polarization profile. The contribution function indicates that the circular
polarization of this \ion{Fe}{2} line forms mainly at heights between 0.3 and 0.5~Mm in the FAL-C model,
which corresponds to the upper photosphere. The $B_{\rm L}$ inferred with the WFA
corresponds to the actual value in the model at a height of 0.2~Mm, just below
the formation region deduced from the contribution function.

The \ion{Fe}{2} line at 280.66~nm, the one at the red edge of the CLASP2 spectral
window, shows a more typical antisymmetric shape with two lobes, with
signals larger than those of the \ion{Fe}{2} line at 279.79~nm. From the
contribution function, the region of formation of the circular polarization appears
to be located at heights between 0.2 and 0.7~Mm in the upper photosphere of the FAL-C model.
The $B_{\rm L}$ inferred with the WFA corresponds to the field expected at those heights.

\begin{figure*}[htp!]
\plotone{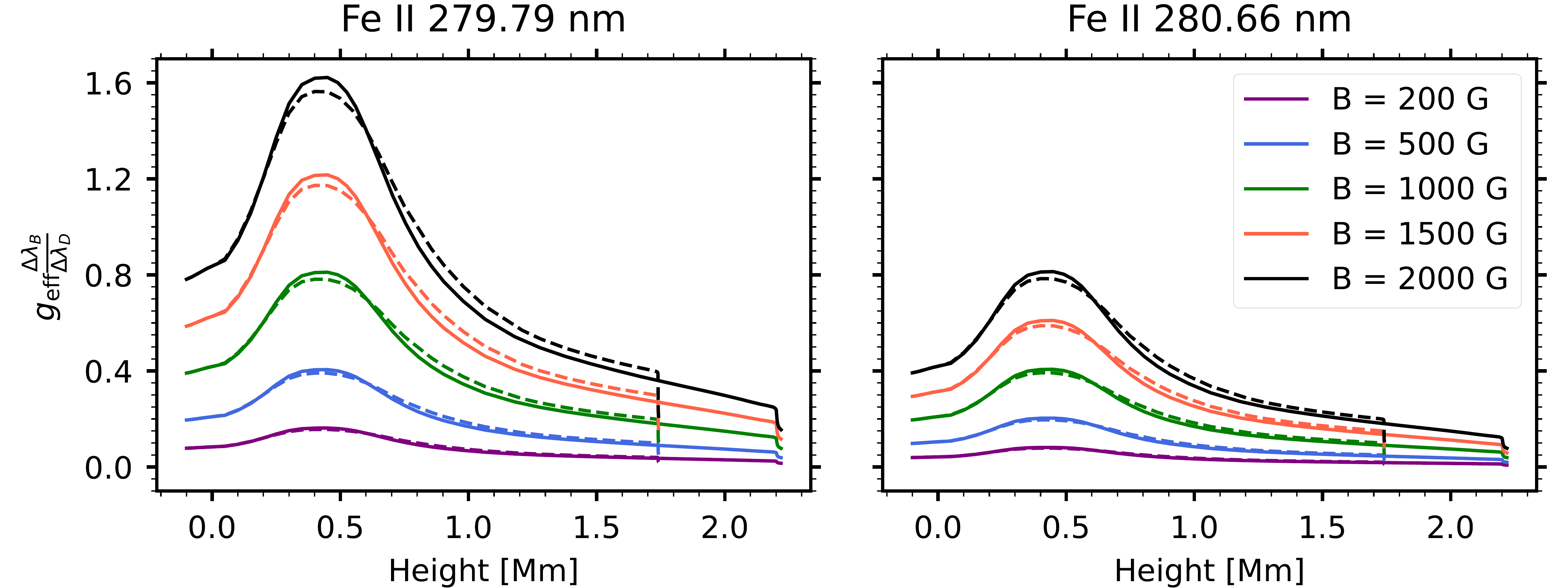} 
\caption{Stratification of the ratio
$g_{\rm eff}\frac{\Delta\lambda_B}{\Delta\lambda_D}$ in the FAL-C (solid curves)
and FAL-P (dashed curves) models, for a uniform magnetic field, for the \ion{Fe}{2}
lines at 279.79~nm (left panel) and at 280.66~nm (right panel). The colors of the
curves indicate the magnetic field strength (see legend in the right panel).  
}
\label{fig:wfaCondition}
\end{figure*}

As illustrated at the beginning of this section, the WFA can only be applied if
Eq.~(\ref{Eq:WFAcondition}) is satisfied. In Fig.~\ref{fig:wfaCondition} we show
the stratification of the ratio
$g_{\rm eff}\frac{\Delta\lambda_B}{\Delta\lambda_D}$ for the \ion{Fe}{2} lines
at 279.79 and 280.66~nm for different uniform magnetic fields, for both the
FAL-C and FAL-P models. 
In typical applications to solar spectropolarimetry, the weak field
condition of the WFA is often satisfied when the ratio in Eq.~\ref{Eq:WFAcondition} is smaller than about 0.5.
Consequently, for these lines that form around the temperature
minimum ($\sim0.5$~Mm, which would be the worst case scenario), the WFA would be
suitable for magnetic fields up to about 500~G for the \ion{Fe}{2} line at
279.79~nm and up to about 1000~G for the \ion{Fe}{2} line at 280.66~nm.

In summary, the \ion{Fe}{2} lines at 279.79 and 280.66~nm 
form in the upper photosphere, considerably deeper than the \ion{Mg}{2} and \ion{Mn}{1} in the same
spectral range, and the WFA can be applied for a quick inference of $B_{\rm L}$ 
for longitudinal components of the magnetic field up 
to 500~G (for the 279.79~nm line) and up to 1000~G (for the 280.66~nm line).

\subsection{Application to the CLASP2 data} \label{sec:Results_data}

The intensity and circular polarization profiles of the \ion{Mg}{2} h and k lines
and of the \ion{Mn}{1} resonant lines observed by CLASP2 in a solar plage region
have been analyzed in \cite{Ishikawa2021a}, mapping the longitudinal component
of the magnetic field from the photosphere
to the upper chromosphere. The $B_{\rm L}$ in the lower chromosphere was inferred by
applying the WFA to the \ion{Mn}{1} resonant lines. The $B_{\rm L}$ in the middle
chromosphere was inferred by applying the WFA to the outer lobes of the
\ion{Mg}{2} h line. The $B_{\rm L}$ in the upper chromosphere was inferred by applying the WFA to the
inner lobes of the \ion{Mg}{2} h and k lines. The determination of the photospheric
$B_{\rm L}$, however, resulted from a Milne-Eddington inversion of the \ion{Fe}{1} lines
at 630.15 and 630.25~nm, from data acquired in a simultaneous observation by the Hinode
SOT-SP \citep{Kosugi2007}. These lines form in the lower layers of the solar
photosphere \citep{Grec2010, Faurobert2012}. In this section we extend the work
of \cite{Ishikawa2021a} by adding the inference of $B_{\rm L}$ by applying the WFA
to the \ion{Fe}{2} lines at 279.79 and 280.66~nm, as we did with the synthetic
profiles in Sec.~\ref{sec:Results_Theor}.

Figure~\ref{fig:wfa} shows the fractional circular polarization in one pixel of the
CLASP2 observation of a plage region. We apply the WFA to the same seven\footnote{Note that 
we distinguish two spectral regions for the \ion{Mg}{2} h
line, and thus there are seven spectral regions in total.} spectral
regions indicated in Sec.~\ref{sec:Results_Theor}.
We get $B_{\rm L}$ between 179 and
193~G in the upper chromosphere (from the inner lobes of the \ion{Mg}{2} h and k
lines), 227~G in the middle chromosphere (from the outer lobes of the \ion{Mg}{2}
h line), and between 370 and 378~G in the lower chromosphere (from the \ion{Mn}{1}
lines). To this $B_{\rm L}$ stratification we can add 646~G in the upper photosphere 
(derived from the \ion{Fe}{2} line at 280.66~nm).
As expected, we obtain a $B_{\rm L}$ which increases with
the depth of the region of formation of each spectral range.

In contrast, we find that the inferred $B_{\rm L}$ is 219~G 
when the WFA is applied to the
\ion{Fe}{2} at 279.79~nm. This is in apparent contradiction with the findings in
Sec.~\ref{sec:Results_Theor}, as this value is very close to the one obtained
from the outer lobes of the \ion{Mg}{2} h line, which forms in the middle
chromosphere whereas, according to the numerical modeling of Sect.\ref{sec:Results_Theor},
the \ion{Fe}{2} 279.79~nm line should be the one that forms in the lowest part of the atmosphere,
among the lines included in this work. This discrepancy can be understood by analyzing the formation of
this spectral line in two semi-empirical models representative of the the quiet Sun
(FAL-C) and a plage region (FAL-P).

\begin{figure*}[htp!]
\centering
\plotone{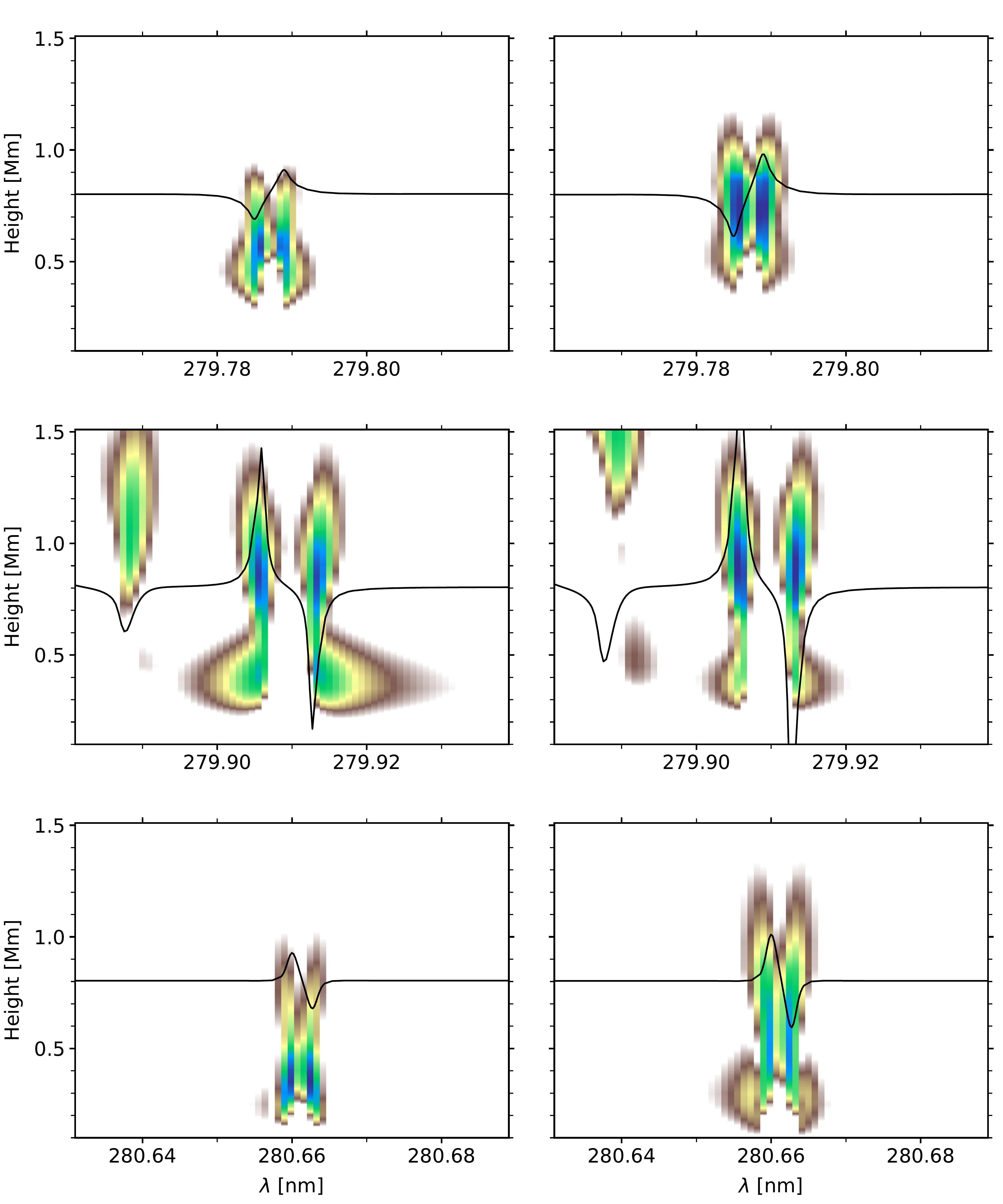}
\caption{Fractional circular polarization $V/I$ calculated in the FAL-C (top row) and FAL-P
(bottom row) semi-empirical model atmospheres. Each column shows the wavelength range around a spectral
line: \ion{Fe}{2} 279.79~nm (left), \ion{Mn}{1} 279.91~nm (middle), and \ion{Fe}{2} 280.66~nm (right).
The colored background shows the absolute value of the logarithm of the normalized contribution
function for the circular polarization, with the blue (brown) showing the regions with larger (lower)
contribution.
}
\label{fig:contr}
\end{figure*}

In Fig.~\ref{fig:contr}, we show the absolute value of the logarithm of the contribution function 
for the two \ion{Fe}{2} (279.79 and 280.66~nm) and the \ion{Mn}{1} (279.91~nm) lines calculated in the FAL-C
and FAL-P model atmospheres. The blue regions correspond to the largest contribution to the circular polarization.
In the quiet Sun model (top row of Fig.~\ref{fig:contr}) the largest contribution for the two
\ion{Fe}{2} spectral lines is located at very similar heights, deeper in the atmosphere compared to
the \ion{Mn}{1} spectral line. However, in the plage region model (bottom row of Fig.~\ref{fig:contr})
the largest contribution for the 279.79~nm line is higher in the atmosphere with respect to the other
spectral lines. In this model this \ion{Fe}{2} line forms at a height which is very similar to 
that of the \ion{Mn}{1} line. Instead, the \ion{Fe}{2} 280.66~nm line formation region is found always
in deeper layers than the \ion{Mn}{1} line.

In summary, while in quiet Sun conditions both \ion{Fe}{2} lines form in relatively similar regions,
in the plage region the formation of the \ion{Fe}{2} line at 279.79~nm is shifted upwards relative to
the other lines. This explains why, when applying the WFA approximation to the CLASP2
data in the plage region, we infer $B_{\rm L}$ values from the \ion{Fe}{2} line at 279.79~nm
that are smaller than those inferred from both the \ion{Fe}{2} line at 280.66~nm and the
\ion{Mn}{1} lines.

\begin{figure*}[htp!]
\plotone{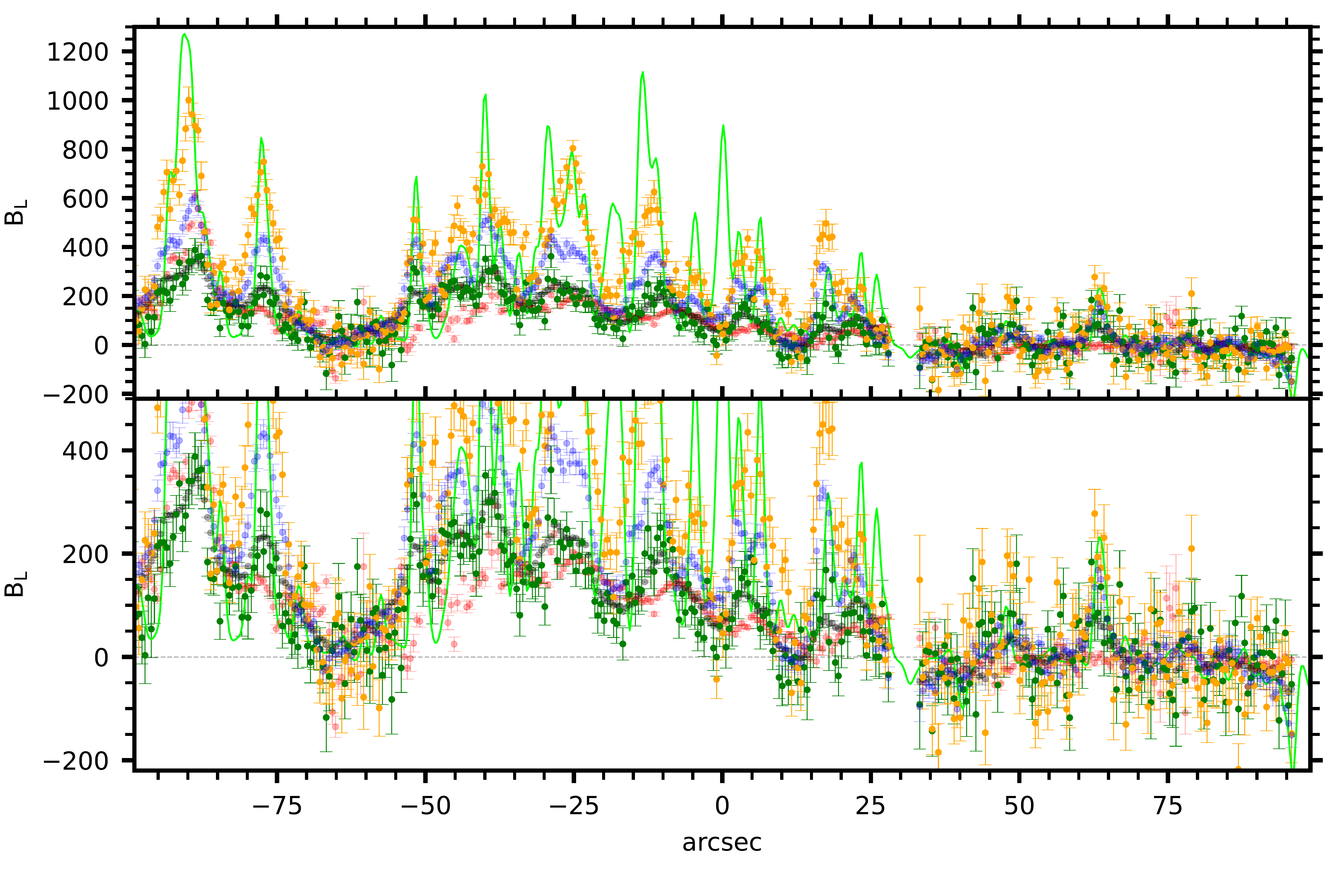} 
\caption{Inferred $B_{\rm L}$ along the CLASP2 spectrograph slit.
The light-green curve corresponds to the lower photosphere, obtained from the
Milne-Eddington inversion of the Hinode/SOT-SP observation. The colored circles
correspond to the inference applying the WFA to the seven spectral regions shown
in Fig.~\ref{fig:wfa_theor}:
red circles: upper chromosphere inferred from the inner lobes of the \ion{Mg}{2}
h and k lines; blue circles, middle chromosphere inferred from the outer lobes
of the \ion{Mg}{2} h line; black circles, bottom chromosphere inferred from the
\ion{Mn}{1} lines; orange circles, upper photosphere inferred from the \ion{Fe}{2} line
at 280.66~nm, and dark green circles, middle chromosphere (top photosphere) in
the plage region (quiet Sun region) inferred from the \ion{Fe}{2} line at 279.79~nm.
The lower panel shows a zoom in the magnetic field axis.
}
\label{fig:slit}
\end{figure*}

With this more complete picture of the region of formation of the circular polarization
of the lines in the CLASP2 spectral window we apply the WFA to the seven spectral
regions specified in Fig.~\ref{fig:wfa_theor} in all pixels of the CLASP2
plage region observation. Fig.~\ref{fig:slit} shows the inferred $B_{\rm L}$
along the spectrograph slit. We include in the figure the $B_{\rm L}$ WFA
inference from the \ion{Mg}{2} h and k lines (top and middle chromosphere, red and black,
respectively) and the \ion{Mn}{1} lines (lower chromosphere, blue
dots), which were presented in \cite{Ishikawa2021a}, as well as the result of
their inversion of the Hinode data (light green curve). The $B_{\rm L}$ values we infer
from the \ion{Mg}{2} and \ion{Mn}{1} spectral lines are very similar to those shown 
in \cite{Ishikawa2021a}.

The important new result in Fig.~\ref{fig:slit} is the $B_{\rm L}$ resulting from the
application of the WFA to the \ion{Fe}{2} lines at 279.79 and 280.66~nm (dark green and
orange dots). The $B_{\rm L}$ inferred from the \ion{Fe}{2} line at 280.66~nm is larger
than for the other lines in the CLASP2 spectral window, and relatively close to the
Milne-Eddington inversion of the Hinode data. This is in agreement with the expected
photospheric region of formation of this spectral line. The $B_{\rm L}$ inferred from the
\ion{Fe}{2} line at 279.79~nm is instead similar to the values inferred for the middle
chromosphere inside the plage region (slit positions 
from -99 to $\sim$~-70~arcsec and from $\sim$~-50 to $\sim$25~arcsec). Instead,
in the nearby quiet Sun the inferred $B_{\rm L}$ is similar to the inference from the
\ion{Fe}{2} line at 280.66~nm, which is fully consistent with the change in the region of
formation shown in Fig.~\ref{fig:contr}.

\section{Conclusions} \label{sec:conclusions}

We investigated theoretically the formation of the intensity and circular polarization
profiles of the \ion{Fe}{2} lines located in the CLASP2 spectral window, comparing
their region of formation with those of the \ion{Mg}{2} and \ion{Mn}{1} lines in the
same spectral window, which were considered in previous studies.
To this end, we solved the RT problem in
a semi-empirical model atmosphere representative of the quiet Sun, to which we added an ad-hoc
stratification of the magnetic field. In particular, we studied the formation of those
\ion{Fe}{2} spectral lines and their suitability to infer $B_{\rm L}$ by applying the WFA.
We demonstrated the validity of this approximation and apply it to the CLASP2 data, extending
the mapping of the longitudinal component of the magnetic field in \cite{Ishikawa2021a} to additional layers of the solar atmosphere, between
the upper photosphere and the lower chromosphere.

We compared the intensity emerging from the FAL-C model atmosphere,
with observations by IRIS and CLASP2, 
for two LOS, one at disk center and one close to the limb.
With the exception of the 279.47 nm line, all the Fe II lines in this spectral region show
good qualitative agreement between the
numerical calculations and the observations. This indicates that the \ion{Fe}{2}
model atom presented in \cite{Afonso2023b}, is
suitable to model the \ion{Fe}{2} lines in this spectral window. In particular, the
model reproduces the peculiar behavior of the \ion{Fe}{2} line at 279.79~nm, almost
absent at disk center, but with a clear emission at the solar limb or in active
regions such as plage regions. We note that there is a \ion{Fe}{2} line at 369.94~nm,
in the red wing of the \ion{Ca}{2} H line, which shows the same behavior, with
an absorption profile at disk center, but an emission profile closer to the limb
\citep{Cram1980, Watanabe1986}. In a future investigation, 
we will study the formation of this spectral line,
which can be observed with ground-based facilities such as the Daniel K. Inouye Solar Telescope \cite[DKIST,][]{DKIST}.

We have found that only two of these five \ion{Fe}{2} lines (279.471, 279.787, 280.012, 280.485, and 280.661~nm) show circular polarization
signals above the noise level in the CLASP2 observations. 
By solving
the RT problem in the FAL-C model (Sec.~\ref{sec:Results_Theor}), 
we showed that both lines form in the photosphere
when the thermodynamical conditions of the observed target 
are similar to those of the quiet Sun.
We showed that the WFA of these lines can be applied for $B_{\rm L}$
up to 500~G for the 279.79~nm line and up to 1000~G for the 280.66~nm line.

We also found that the exact
region of formation of the \ion{Fe}{2} line at 279.79~nm
depends on the particular physical conditions of the atmospheric plasma. While in the
typical conditions of the quiet Sun this line forms in the photosphere, like the \ion{Fe}{2}
line at 280.66~nm, for the typical conditions of a solar plage the region of formation
is shifted upward, closer to the \ion{Mg}{2} h and k line wings, even above the
\ion{Mn}{1} lines which form in the lower chromosphere. This difference in height of
formation explains the apparent discrepancy in the inferred $B_{\rm L}$
which, in the plage region, is smaller for the \ion{Fe}{2} line at 279.79~nm than for
both the \ion{Fe}{2} line at 280.66~nm and the \ion{Mn}{1} lines. 
The \ion{Fe}{2} line at 280.66~nm appears instead to be consistently
photospheric. It has a formation region around the temperature minimum, and the inferred
$B_{\rm L}$ is close to that shown in \cite{Ishikawa2021a} obtained with a
Milne-Eddington inversion of simultaneous observations with Hinode/SOT-SP of the
\ion{Fe}{1} lines at 630.15 and 630.25~nm.

It is important to emphasize that there are additional spectral lines in the
CLASP2 spectral window, from species other than \ion{Fe}{2}, which show significant
circular polarization signals. Our ongoing studies of these signals should help
complete the mapping of $B_{\rm L}$ from the upper photosphere to the top of the chromosphere.

The results of this work demonstrate the suitability of the \ion{Fe}{2} lines at
279.79 and 280.66~nm to supplement the information 
provided by the \ion{Mg}{2} and \ion{Mn}{1} lines of the CLASP2
spectral window to map $B_{\rm L}$ across the solar atmosphere
\citep[see][]{Ishikawa2021a}. 
Therefore, we conclude that it is possible to retrieve information about $B_{\rm L}$ 
from the photosphere to the upper chromosphere just below the transition region from
spectropolarimetric observations in the spectral window of the \ion{Mg}{2} h and k lines.
This finding strongly emphasizes the importance and timeliness for deploying a new space mission equipped with CLASP2-like capabilities, as this would likely lead to a breakthrough in our ability to probe 
the magnetism of the solar atmosphere from the upper photosphere to the base of the corona.

While our conclusions strongly emphasizes the unique diagnostic potential of this narrow near-UV spectral region, future investigations targeting the mapping of the magnetic field throughout the solar atmosphere will obviously benefit from complementary observations from multiple instruments, both space-borne and ground-based, of chromospheric diagnostics in the UV, visible, and infrared solar spectrum.

\acknowledgements

We thank Ryohko Ishikawa (NAOJ) and Hao Li (IAC) for useful discussions on the 
CLASP2 Stokes signals and their interpretation. 
We are grateful to the referee for suggestions that have improved
the presentation of this paper.
We acknowledge the funding received from the European Research Council (ERC)
under the European Union's Horizon 2020 research and innovation programme (ERC
Advanced Grant agreement No 742265). T.P.A.'s participation in the publication is
part of the Project RYC2021-034006-I, funded by MICIN/AEI/10.13039/501100011033,
and the European Union “NextGenerationEU”/RTRP. 
CLASP2 is an international partnership between NASA/MSFC, NAOJ, JAXA, IAC, and IAS; 
additional partners include ASCR, IRSOL, LMSAL, and the University of Oslo.
The Japanese participation was funded by JAXA as a Small Mission-of-Opportunity Program, 
JSPS KAKENHI Grant numbers JP25220703 and JP16H03963, 2015 ISAS Grant for Promoting 
International Mission Collaboration, and by 2016 NAOJ Grant for Development Collaboration. 
The USA participation was funded by NASA Award 16-HTIDS16\_2-0027. 
The Spanish participation was funded by the European Research Council (ERC) under the 
European Union's Horizon 2020 research and innovation programme 
(Advanced Grant agreement No. 742265). The French hardware participation 
was funded by CNES funds CLASP2-13616A and 13617A.


\bibliography{feii}
\bibliographystyle{aasjournal}

\end{document}